\documentclass{article}
\usepackage{arxiv}
\usepackage[utf8]{inputenc} 
\usepackage[T1]{fontenc}  
\usepackage{hyperref}       
\usepackage{url}            
\usepackage{booktabs}       
\usepackage{amsfonts}       
\usepackage{nicefrac}       
\usepackage{microtype}      
\usepackage{graphicx}
\usepackage{natbib}
\usepackage{amsmath}
\usepackage{float}
\usepackage[flushleft]{threeparttable}
\usepackage{graphicx,caption}
\usepackage{amsfonts}
\usepackage{longtable}
\usepackage{bm}
\usepackage{array}

\newcommand{\argmax}{\arg\!\max}

\newcommand*{\tran}{^{\mkern-1.5mu\mathsf{T}}}

\usepackage{scalerel,stackengine}
\stackMath
\usepackage{soul}

\title{A Single Index Model for Longitudinal Outcomes to Optimize Individual Treatment Decision Rules}

\author{ \href{https://orcid.org/0000-0001-5935-6018}{\includegraphics[scale=0.06]{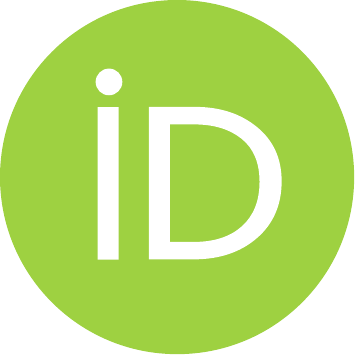}\hspace{1mm}Lanqiu Yao}
\thanks{This work was supported by the National Institute of Mental Health (NIMH) grant 5 R01 MH099003. The authors appreciate the help and support from Drs. Hyung Park, Mengling Liu,  Samprit Banerjee,  Binhuan Wang, and  Jiyuan Hu.
} \\
	Department of Population Health\\
 NYU Grossman School of Medicine\\
	New York, NY 10016 \\
	\texttt{Lanqiu.Yao@nyulangone.org} \\
	\And
\href{https://orcid.org/0000-0002-3964-5899}{\includegraphics[scale=0.06]{orcid.pdf}\hspace{1mm}Thaddeus Tarpey} \\
		Department of Population Health\\
 NYU Grossman School of Medicine\\
	New York, NY 10016 \\
	\texttt{Thaddeus.Tarpey@nyulangone.org} \\
}

\renewcommand{\shorttitle}{\textit{arXiv} Template}

\hypersetup{
pdftitle={A template for the arxiv style},
pdfsubject={q-bio.NC, q-bio.QM},
pdfauthor={Lanqiu Yao, Thaddeus Tarpey},
pdfkeywords={Precision Medicine, Kullback-Leibler Divergence, Prediction Model, Mental Health},
}

\begin{document}
\maketitle

\begin{abstract}

A pressing challenge in medical research is to identify optimal treatments for individual patients. 
        This is particularly challenging in mental health settings where mean responses are often similar across multiple treatments.
        For example,  the mean longitudinal trajectories for patients treated with an active drug and placebo may be very similar but different treatments may exhibit distinctly different individual trajectory shapes.
        Most precision medicine approaches using longitudinal data often ignore information from the longitudinal data structure.
        This paper investigates a powerful precision medicine approach by examining the impact of baseline covariates on longitudinal outcome trajectories to guide treatment decisions instead of traditional scalar outcome measures derived from longitudinal data, such as a change score. 
        We introduce a method of estimating ``biosignatures'' defined as linear combinations of baseline characteristics (i.e., a single index) that optimally separate longitudinal trajectories among different treatment groups. 
        The criterion used is to maximize the Kullback-Leibler Divergence between different treatment outcome distributions. 
        The approach is illustrated via simulation studies and a depression clinical trial. The approach is also contrasted with more traditional methods and compares performance in the presence of missing data.
        
\end{abstract}

\keywords{Precision Medicine \and Kullback-Leibler Divergence \and Prediction Model\and  Mental Health}

\section{Introduction}\label{sec1}
      
      A major goal of precision medicine is to determine optimal treatment decisions rules for individual patients, instead of using the usual ``one-size-fits-all'' approach assuming one therapy works better than the others among all patients~\citep{kosorok2019precision}.
      The traditional strategy of treating all patients with the same treatment is often suboptimal since patients are often highly heterogeneous in many aspects, such as their clinical, genetic, environmental, and social characteristics and hence, there may not be a treatment option that is universally optimal for all patients.
      For example, there is substantial evidence of heterogeneity in treatment response among cancer patients~\citep{dagogo2018tumour}.
      Patients with advanced colorectal or breast cancer might not always benefit from molecular target treatment and comprehensive molecular aberrations studies are required ~\citep{dienstmann2012molecular, de2013molecular}. 
      Heterogeneous treatment response is also common in depression studies.
      For example, only about $30\%$ of patients with depression achieve remission after a single acute phase of treatment~\citep{trivedi2006evaluation}. 
      This heterogeneity presents a significant challenge in medical research, making it difficult to find therapies that are effective in a majority of patients.
      Precision medicine is therefore a promising approach to solving this heterogeneity problem whereby treatment decisions are tailored to individual patients based on a patient's characteristics~\citep{petkova2017generated}.

Numerous methods have been developed to estimate \textit{treatment decision rules} (TDRs) using baseline patient characteristics. 
For example, ~\cite{watkins1989learning} proposed Q-learning, where ``Q'' stands for ``quality''.
It is a reinforcement learning method that estimates the Q-function, which is defined as the conditional expectation of the sum of the current and future outcomes given that treatment decisions are made at all future stages based on the specific TDR.
After~\cite{murphy2007methodological} applied Q-learning in the clinical research area, 
it has been widely studied to optimize TDRs, and the Q-functions can be modeled parametrically, semiparametrically and nonparametrically~\citep{zhao2009reinforcement}.
~\cite{murphy2003optimal} introduced a similar method known as Advantage learning (A-learning). It models a regret function that measures the loss caused by not following the optimal treatment regime at each stage and can be more robust to model misspecification than Q-learning~\citep{schulte2014q}. 
Outcome weighted learning (OWL) is another widely used approach proposed by~\cite{zhao2012estimating}, where a weighted classification function calculated with baseline characteristics is applied to optimize the estimation of treatment rules and the computational problem is solved using support vector machines (SVM).  
Researchers have also utilized  deep learning approaches on precision medicine, such as neural networks and evolutionary algorithms~\citep{uddin2019artificial}. 
However, these machine learning and deep learning methods are very computationally burdensome and the results from these methods are often difficult to interpret.

Regression models have been investigated to develop TDRs \citep[e.g.,][]{petkova2020optimising, park2020single, tian2014simple,henderson2010regret}.
Regression approaches have the advantage that  the resulting TDRs are easy to use and easily interpretable. Additionally, TDRs based on regression modeling are often very competitive compared to more complex machine learning approaches.
\citet{petkova2017generated} proposed a model that uses a linear combination or an \textit{single index} of patients' baseline characteristics, termed a \textit{generated effect modifier} (GEM) or ``biosignature" to optimize treatment decisions. The linear combination defining the biosignature in the GEM is determined by maximizing the interaction effect (or modifying effect) between treatment arms.
\citet{park2020single} introduced a robust \textit{single-index model with multiple links} (SIMML) functions, a generalization of the GEM, in which various treatment-specific nonparametric link functions were used to relate a single linear combination of baseline characteristics to outcomes.

Most medical studies have durations that last weeks and measures are typically collected longitudinally over the course of the study on participants.  A shortcoming in the TDR literature is that the information available from the longitudinal assessments is typically ignored and only the baseline and last measurement is used to define an outcome, e.g., a change score, or controlling for baseline in an analysis of covariance (ANCOVA) model.
TDR's that ignore this temporal longitudinal aspect of the data may be missing important information that could be used to optimize the TDR.  For example, one treatment may have a much quicker onset of action than another but the duration of the specific action of the treatment may wane quicker. 
In such a situation, the two treatments may look similar based on a simple change score outcome, but the distribution of trajectories for the two treatments may be quite distinct.
An additional challenge in medical studies is that they are almost always plagued by missing data.
Similar to the longitudinal structures that are often ignored,  few methods precision medicine approaches account for missing data.
Methods that are robust to missing data are potentially more powerful for developing well-performing TDRs.  

Our objective is to create an effective technique for estimating TDRs that takes use of the data's longitudinal structure and is robust when there is missing data.
The rest of the paper is organized as follows:
in Section \ref{embarcSection} we introduce our motivating example, the EMBARC trial;
Section \ref{lsi} describes the algorithm we developed to identify an optimal linear combination of baseline covariates for treatment decisions; Section \ref{geometric} provides a geometric illustration of the proposed TDR approach;
to evaluate our method,  {a} simulation study is reported in Section \ref{sim}; 
Section \ref{sec:embarc} illustrates an application of our approach with the EMBARC example described in Section \ref{embarcSection}; a brief discussion is provided in Section \ref{disc}.

\section{Motivating example: the EMBARC trial}\label{embarcSection}

Major Depressive Disorder (MDD) is a chronic disease that persistently impacts people's moods and behaviors. 
MDD is highly prevalent in the USA and current studies suggest that approximately 10–20$\%$ of the U.S. population will develop a depressive disorder in their lifetime~\citep{richards2011prevalence}.
The antidepressants, such as Serotonin and noradrenaline reuptake inhibitors (SNRIs), were believed to work in 70$\%$ MDD patients~\citep{dinoff2020meta}. 
However, researchers found the effects of antidepressants were overestimated, i.e., the magnitude of symptom reduction with antidepressants was about only 40$\%$~\citep{khan2012systematic}. The resistance to medication is usually explained by the extreme heterogeneity. 
The clinical trial, Establishing Moderators and Biosignatures of Antidepressant Response in Clinical Care (EMBARC) 
~\citet{trivedi2016establishing} was designed to discover biomarkers for depression treatment response using a large collection of baseline modalities, such as clinical measures, behavior tests, brain imaging modalities. 
The primary goal was to discover biomarkers from this rich set of baseline data to deal with the heterogeneity and develop personalized treatment decision rules that use this rich source of information. 

\begin{figure}
\caption{\textbf{Outcome Trajectories Plots for Participants in the EMBARC Study}}
\centering
\includegraphics[width=0.6\textwidth]{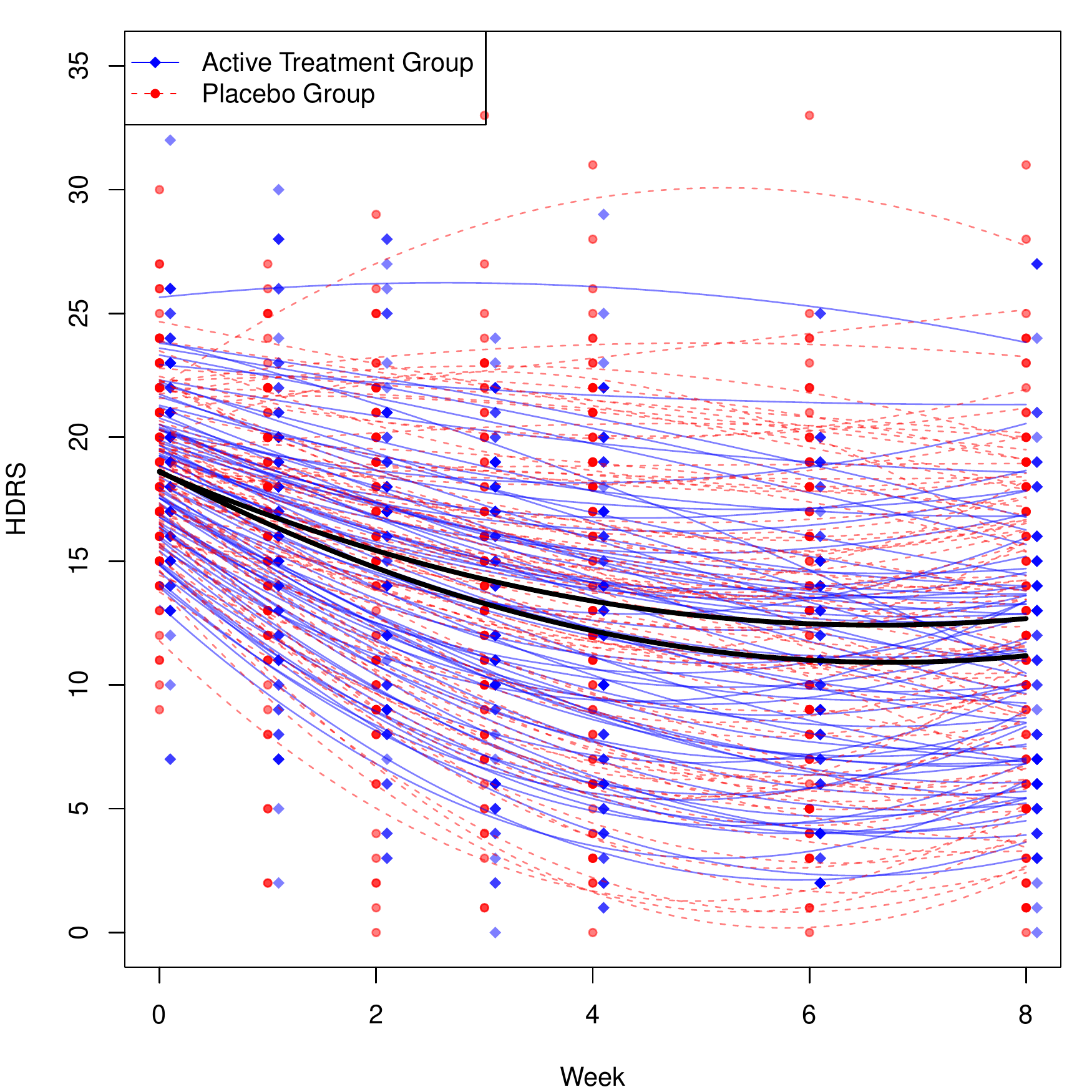}
\label{fig:fig1}
\caption*{
\footnotesize{\textbf{Figure \ref{fig:fig1}:} 
The HDRS scores for the active treatment group (blue diamond points) and the placebo group (red circle points) at corresponding weeks. 
The data was analyzed using mixed effect models with quadratic effects of observation time. 
The blue solid and red dashed curves represent the treatment and placebo groups' estimated outcome trajectories, respectively. 
The average trajectories for the two groups are shown by the two bold black curves (the active treatment mean curve is the lower curve).}}
\end{figure}

The EMBARC study was a multi-site randomized clinical trial with 296 participants recruited from four study sites~\citep{petkova2017statistical}.
Participants were randomly assigned to receive either sertraline or placebo.
The participants were followed up for 8 weeks and the primary outcome, Hamilton Depression Rating Scale (HDRS), was evaluated for each participant at each follow-up time. 
The HDRS is a measure for the severity of depression with lower scores indicating less severe depression.
Quadratic mixed-effects models with linear and quadratic terms of time (weeks) were fit to the longitudinal data. The estimated outcome trajectories versus time are illustrated in Figure \ref{fig:fig1}.
The blue solid and red dashed curves represent the estimated trajectories for the active treatment and placebo group, respectively. 
The blue and red dots show the observed HDRS values for individual patients. 
The mean outcome trajectories are represented by the thick black curves (the mean trajectory for the active treatment group is the lower one). 
Note that the average trajectories for drug and placebo are very similar to one another.

The baseline measures that were collected in the EMBARC study include demographics (e.g., age and gender), clinical measurements (e.g., anxiety level, ranger attacks, hypersomnia, and fatigue), and cerebral cortical thickness data from MRI, task-based functional MRI (fMRI), {and EEG from scalp electrodes}.
It is quite common in depression studies that no single baseline measure will explain a substantial portion of variability in the outcome measures, but instead, each baseline measure may contribute some minimal amount of information.  
Therefore, a driving goal of precision medicine is to discover powerful biosignatures for  defining TDRs that combine information across multiple baseline features.

\section{A single-index model to optimize Kullback-Leibler Divergence}\label{lsi}

The primary idea for this paper is to develop a linear biosignature (i.e., a linear combination of baseline features) that optimally separates treatment group trajectories in terms of Kullback-Leibler Divergence in order to develop strong performing TDRs.
This work  focuses on randomized clinical trials. 
Let $y_{ik}(t)$ denote the functional  (e.g., longitudinal trajectory) outcome for subject  $i \in \{1,..., n_k\}$ 
in the $k$th treatment group  where $n_k$ is the number of participants for the $k$th group; $k= 1,.., K$ and $K$ is the number of intervention arms. 
In a longitudinal study, $y_{ik}(t)$ is only observed at a finite number of design time points (possibly with error). 
Denote $\tilde{y}_{ijk}$  as the observed outcome measured for the given subject at time $t_{ijk}$, where $j \in \{1,..., m_{ik}\}$ is the index of observations and let $\tilde{\bm Y}_{ik} = (\tilde{y}_{i1k}, \tilde{y}_{i2k}, ..., \tilde{y}_{im_{ik}k})\tran$ denote the vector of outcomes for the $i$th subject in the $k$th group.
In a balanced longitudinal study with no missing values, the measurement times will usually follow a protocol with a common set of follow-up times, i.e., $t_{ijk} = t^*_j$ {with} a total number of $m$ outcomes collected {per subject}.
Denote the vector of observed times for subject $i$ as $\bm t_{ik} = (t_{i1k}, t_{i2k}, ..., t_{im_{ik}k})\tran$, where $m_{ik} \leq m$. That is, a subject may have missing time points or drop out of the study, the total observed time is less or equal to the total designed follow-up time points, $m$. 
Let $\bm x_{ik}\tran= (x_{ik1}, ..., x_{ikp})\in \mathbb{R}^p$ 
denote the set of baseline covariates. 
The TDRs will be developed for a linear biosignature defined as a linear combination of baseline covariates:
 $\bm \alpha\tran \bm x_{ik}$. 
To ensure identifiability of the model, we impose the constraint $||\bm \alpha||_2 = 1$.
To incorporate a subject's baseline characteristics into a longitudinal model, we consider the following mixed-effects model for $\tilde{\bm Y}_{ik}$:
\begin{equation}\label{lme}
    \tilde{\bm Y}_{ik} = \bm G(\bm t_{ik}) \big(\bm \beta_k + \bm b_{ik} + \bm \Gamma_k (\bm \alpha \tran \bm x_{ik}) \big) + \bm \epsilon_{ik},
\end{equation}
with the biosignatrue $\bm \alpha \tran \bm x_{ik}$  and  $\bm \Gamma_k$ is its {fixed-effect} coefficient vector with dimension $q \times 1$;
$\bm \beta_k$ is also a $q\times 1$ coefficient of fixed effects, which {represents} the main effects in the $k$th group; 
$\bm b_{ik} \sim N(\bm 0, \bm D_k)$ is the subject-specific random effects, assumed to be multivariate normal; 
$\bm \epsilon_{ik}\sim N(0, \sigma_k^2 \bm I)$ is the vector of random errors, which is assumed independent of the random effects; 
$\bm G(\bm t_{ik})$ is the design matrix with dimension $m_{ik} \times q$, which is constructed with a set of suitable basis functions of the observation times $t_{i1k}, t_{i2k},..., t_{im_{ik}k}$, i.e. $\bm G(\bm t_{ik}) = \big(\bm g(t_{i1k}), \bm g(t_{i2k}),..., \bm g(t_{im_{ik}k})\big) \tran$, where $\bm g(t) = (g_1(t), g_2(t), ..., g_q(t))\tran$ is a $q \times 1$ sequence of basis functions for time $t$.
Thus the biosignature is embedded in a longitudinal model and has an interaction with the observation time. 

\subsection{Purity function}\label{purity}

TDRs are generally based on combinations of baseline features that have treatment modifying impacts on the outcome and in standard regression settings, this often corresponds to finding treatment effect modifiers that maximize interaction effects.
Correspondingly, in order to optimize a TDR based on the functional nature of the outcome, an optimal choice of $\bm{\alpha}$ in (\ref{lme}) is determined in order to maximally distinguish the treatment groups in terms of the overall trajectories shapes.
Let $\bm z_{ik}$ denote the coefficients for the trajectory of subject $i$ in the $k$th group from (\ref{lme}), that is, 
\begin{equation}\label{coef0}
\bm z_{ik} = \bm \beta_k + \bm b_{ik} + \bm \Gamma_k (\bm \alpha \tran \bm x_{ik}).
\end{equation}
Conditional on a given biosignature $\bm \alpha \tran \bm x$, from the model assumptions, the  coefficient distribution (\ref{coef0}) is multivariate normal: 
\begin{equation}
    \bm z_{ik} | \bm \alpha \tran \bm x \sim MVN \Big( \bm \beta_k + \bm \alpha \tran \bm x, \bm D_k \Big). 
\end{equation}
To measure the differences among the trajectory shape distributions as determined by the coefficients from different groups, the Kullback-Leibler Divergence (KLD) is applied, which measures {the degree of divergence} of one probability distribution $F_1$ from another probability distribution $F_2$:
\begin{equation} \label{ch2.2}
D_{KL}(F_1 || F_2) = \int_{-\infty}^{+\infty} f_1(x) \log(\frac{f_1(x)}{f_2(x)})dx,
\end{equation}
where $f_1$ and $f_2$ denote the probability density functions of $F_1$ and $F_2$, respectively. Larger KLD between distributions corresponds to higher \textit{purity}, i.e., the distributions overlap less. 
Note that $D_{KL}(F_1 || F_2)$ is nonnegative.
Here, we focus the scenario with two different treatment groups, i.e. $k \in \{1,2\}$.
The \textit{purity} function of the biosignature $\bm \alpha \tran \bm x$ is then defined as:
\begin{equation} 
q(\bm \alpha \tran \bm x) =  D_{KL}(F_1 || F_2)(\bm \alpha \tran \bm x) + D_{KL}(F_2 || F_1)(\bm \alpha \tran \bm x),
\end{equation}
which represents the KLD between the distribution of group 1 ($F_1$) and the distribution of group 2 ($F_2$), given the biosignature $\bm \alpha \tran \bm x$. 
The function $q(\cdot)$ is a subject-level purity measure function since it depends on the subject-specific baseline covariates $\bm x$. 
In order to obtain a population-level of purity, we integrate over the entire population: 
\begin{equation}
    Q(\bm \alpha) = \int q(s) f_{\alpha} (s)ds, 
\end{equation}
where $f_\alpha(\cdot)$ is the probability density function (PDF) for the distribution of biosignature $\bm \alpha \tran \bm x$ (which will be identical across treatment groups in a randomized trial). We note the dependence of the biosignature PDF on ${\bm \alpha}$ by $f_\alpha$. 
 A larger purity, $Q(\bm \alpha)$, represents a larger difference in outcomes over time between the two treatment populations.
A larger difference between interventions indicates that it is easier to make a treatment decision, since one treatment is significantly different and superior than the others conditional on a set of baseline measures. 
Therefore, identifying a linear combination of baseline covariates to achieve the highest purity will lead to an optimal separation of the outcome distributions, which in turn will allow the estimation of {well-performing} TDRs. 
Given the mixed-effect model that incorporates the biosignature in (\ref{lme}), from the supporting information, the purity function $Q(\cdot)$ can be expressed as:
\begin{equation} \label{pur}
    Q(\bm \alpha) = a_1 + a_2 \bm \mu_x \tran \bm \alpha + a_3 \bm \alpha \tran (\bm \mu_x \bm \mu_x \tran + \bm \Sigma_x) \bm \alpha, 
\end{equation}
where 
\begin{equation}\label{aaa}
    \begin{aligned}
        a_1(\bm{\alpha} )= & -p + \frac{1}{2}\text{tr}(\bm D_1^{-1}\bm D_2) +  \frac{1}{2}\text{tr}(\bm D_2^{-1}\bm D_1) + \frac{1}{2} (\bm \beta_1 \tran - \bm \beta_2 \tran) (\bm D_1^{-1} + \bm D_2^{-1}) (\bm \beta_1 - \bm \beta_2),\\
        a_2(\bm{\alpha} ) = & (\bm \Gamma_1 \tran - \bm \Gamma_2 \tran)(\bm D_1^{-1} + \bm D_2^{-1}) (\bm \beta_1 - \bm \beta_2),\\
        a_3(\bm{\alpha} ) = & \frac{1}{2}  (\bm \Gamma_1 \tran - \bm \Gamma_2 \tran)(\bm D_1^{-1} + \bm D_2^{-1}) (\bm \Gamma_1 - \bm \Gamma_2). \\ 
    \end{aligned}
\end{equation}
That is,
$a_1, a_2$ and $a_3$ are scalars {defined in terms of} $\bm \beta_k, \bm \Gamma_k, {\bm D}_k, k = 1,2$.
Note that by varying ${\bm \alpha}$, the mixed-effect parameter values in these equations will vary as well, e.g., we could write,  for example, ${\bm D}_1({\bm \alpha)}$ instead of $\bm{D}_1$ but we suppress this dependency here for notational convenience.

\subsection{Optimization and estimation of $\hat{\bm \alpha}$}\label{opt_est}

In this section, we describe the approach for estimating the model parameters, including the biosignature coefficients ${\bm \alpha}$, in order to maximize the KLD.
Since $\bm{x}$ are baseline covariates, the estimation of the model parameters describing the distribution of $\bm{x}$, can be estimated independent of $\bm \alpha$. 
We will use a working approximation model of $\bm{x} \sim N(\bm{\mu}_x, {\bm \Sigma}_x)$ in our derivations to follow.
As noted above, the estimators $\hat{\bm \beta}_k, \hat{\bm \Gamma}_k$ and $\hat{\bm D}_k$ from (\ref{lme}) will depend on $\bm{\alpha}$.  
In general, closed form expressions for the parameter estimates do not exist and we will use numerical methods to simultaneously maximize the likelihood for the mixed-effect model parameters in (\ref{lme})
(with a set of basis functions ${\bm g}(t)$) and estimating $\bm{\alpha}$ to maximize the KLD to maximize (\ref{pur}).
Therefore, the $\bm \alpha$ that maximizes the separations (i.e., KLD) between distributions (e.g., active treatment verse placebo) is assessed from: 
\begin{equation}\label{estpur2}
    \hat{\bm \alpha}^{\text{KLD}} =   \argmax_{\bm \alpha} \text{ } \hat{\bm Q} (\bm \alpha),
\end{equation}
where parameters from the mixed-effect model in $\bm{Q}$ are determined by maximizing the likelihood function.
Given a set of baseline covariates $\bm x$ with dimension $p$ and a selection of {basis functions ${\bm g}(t)$}, we apply Nelder-Mead algorithm~\citet{nelder1965simplex} to determine parameter estimates that numerically optimize the purity function. 
That is, $\hat{\bm \alpha}^{\text{KLD}}$ is estimated using the Nelder-Mead algorithm on (\ref{estpur2}) and then is set to have unit $L_2$ norm (e.g. $||\hat{\bm \alpha}^{\text{KLD}}||_2$ = 1).

\subsection{Treatment decision rule}\label{tdrsec}

A treatment decision rule is defined as a function $\mathcal{D}(\bm x)$ that maps a subject's baseline characteristics $\bm x$ to one of the $K$ treatment options. 
For illustration, here we consider the case of $K=2$ treatment options, coded as $1$ and $2$.
In order to make a TDR for a longitudinal data, we will extract a scalar summary from the outcome trajectories (curves of outcome vs observation time), since the curves do not have nature ordering. 
The most commonly used summary measure is the ``change score", which is calculated as the differences between the last observation as the first measure.
However, it only uses two outcomes for each subject and hence does not utilize all the information available in the data.
It also ignores the information on the shape of the trajectory.
In order to extract a meaningful scalar measurement that summarizes a function $f(\cdot)$,  ~\cite{tarpey2021extracting} considered the average tangent slope (ATS), which is the average of the tangent slope at each time point $t$ on the trajectory interval. 
Since the slope of the tangent line is the derivative of the function at time $t$, the ATS extracts the average rate of change (e.g., improvement) across the study time interval.
The ATS is given by
      \begin{equation}\label{ats00}
      \textbf{Average Tangent Slope (ATS): } \text{ }
      \text{ } \frac{1}{b-a} \int_{a}^b f^\prime (t) dt= \frac{ f(b) - f (a)}{b-a},
      \end{equation}
where $a$ and $b$ are the start and end time in the study, respectively. 
That is, the ATS is the average of the instantaneous rates of improvement across the time domain.
The ATS is calculated from the estimated fixed-effects and hence gains efficiency by incorporating information from all the data. 

      The TDR of our single-index model based on maximizing KLD is then made by using the ATS criterion. Assume a larger value is preferred, i.e., a larger rate of improvement across time corresponds to a better outcome.
      After the estimation of $\bm \alpha$, the TDR is then defined as:
        \begin{equation}\label{tdr}
      \mathcal{D}(\bm x) = I \Big ( \frac{\bm g(t_m^*)\tran - \bm g(t_1^*)\tran}{t_m^* - t_1^*} \big(\hat{\bm \beta}_2 + \hat{\bm \Gamma}_2 (\hat{\bm \alpha} \tran \bm x )\big) > \frac{\bm g(t_m^*)\tran - \bm g(t_1^*)\tran}{t_m^* - t_1^*} \big(\hat{\bm \beta}_1 + \hat{\bm \Gamma}_1 (\hat{\bm \alpha} \tran \bm x) \big)
                                    \Big) + 1,
      \end{equation}
      where $t_1^*$ and $t_m^*$ are the first and last designed outcome measure time, respectively; 
      $\bm g(t) \in \mathbb{R}^{q\times 1}$ is a sequence of basis function evaluated at time $t$. 
      ~\cite{tarpey2021extracting} also demonstrated the properties of the ATS estimator via simulation studies under missing data scenarios.
      If a subject is missing their last scheduled assessment, the change score method of estimation can produced seriously biased estimates of treatment effects when the outcome trajectories are nonlinear.  
      However, the TDR based on the KLD uses parameter estimates for the full data in (\ref{tdr}) and hence suffers far less from the loss of efficiency or bias that hinders the change score method.
      
      \section{Geometric illustrations with quadratic trajectories}\label{geometric}
      
      In this section, we illustrate geometrically how the KLD varies as a function of the biosignature $\bm \alpha \tran \bm x$ using quadratic trajectories as an example. 
      Many randomized clinical trials (RCTs) are of fairly short duration (e.g., 6-12 weeks).  
      For many treatment studies (e.g., for pain and depression), it is common to see some subjects improve immediately (e.g., due to a placebo effect) followed by a deterioration while others may exhibit {delayed response} to therapy  but later improve; on the other hand, some subjects exhibit a linear trend in improvement.  
      These major characteristics of the therapy effects indicate that quadratic curves will {often} provide a good fit to the observed data and this motivates our use of quadratic curves for illustration.
      The coefficient vector in (\ref{coef0}), $\bm z_{ik} = \bm \beta_k + \bm b_{ik} + \bm \Gamma_k \bm \alpha \tran \bm x_{ik}$, is then a $3 \times 1$ vector consisting of an intercept, a coefficient for the linear term (``slope'') and a coefficient for the quadratic term (``concavity''). 
      In a randomized trial, the intercept term represents the average outocme at baseline and should be the same across treatment groups. 
      Also, the intercept provides no direct information on the shape of the trajectory. 
      Thus, for our illustration, we focus on the shape information only, i.e., the slope and concavity values to represent the curves which can be graphically depicted in a 2-dimensional coordinate system ($x$-axis for slope and $y$-axis for concavity). The bivariate normal distribution for these coefficients will be illustrated using elliptical contours of equal density. 
      The center of the contour ellipses will represent the means of the distribution given by the second and third elements of the fixed-effect coefficient vectors $\bm \beta_k$'s for groups $k=1,2$.
The length of the major and minor axis of the ellipses are proportional to the eigenvalues of $\bm D_k$, the covariance matrices of the random effects in $\bm z_{ik}$. 

\begin{figure}[H]
\centering
\caption{\textbf{Illustration of Contour Plots for Trajectories Separations}}
\includegraphics[width=0.8\textwidth]{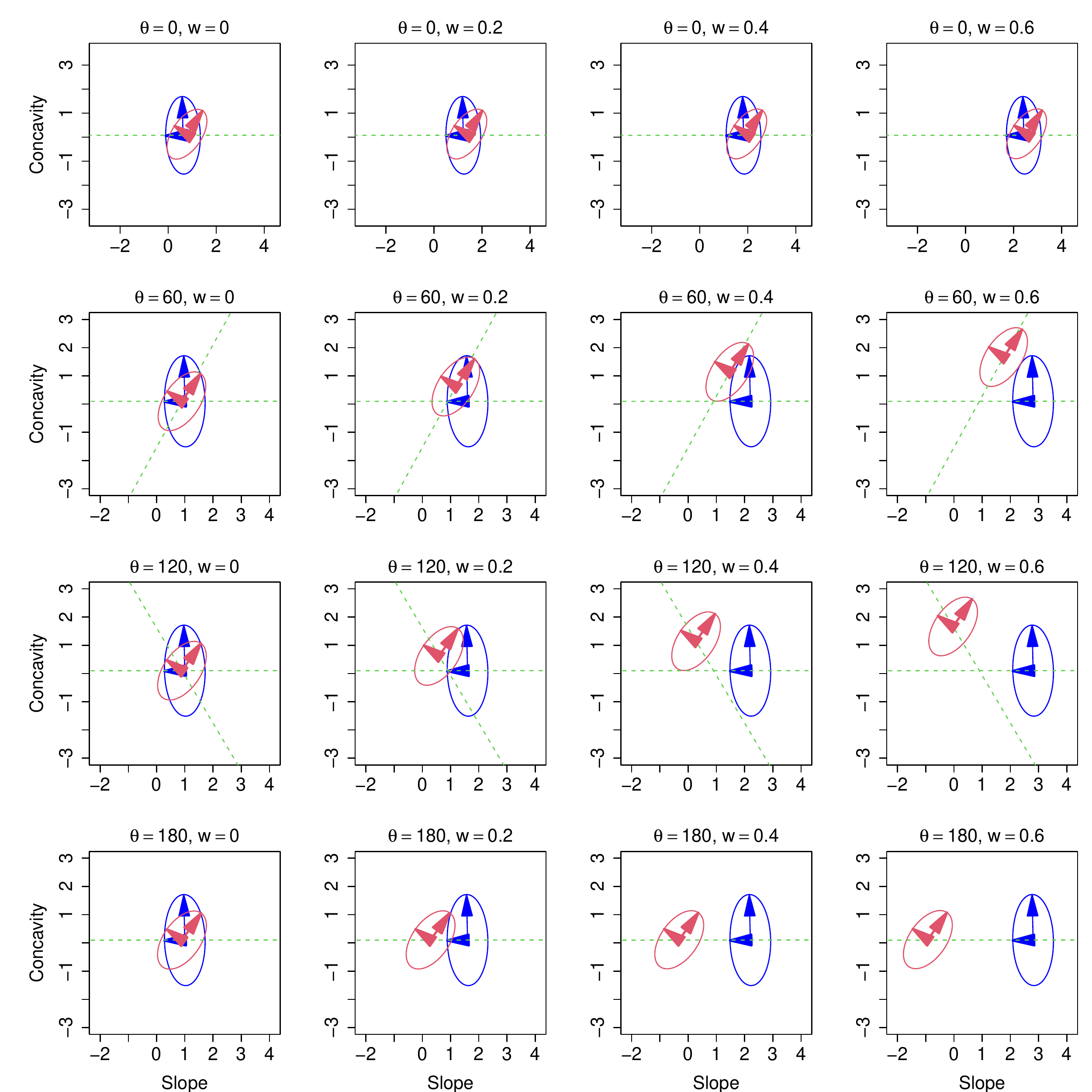}
\caption*{
\footnotesize{\textbf{Figure \ref{fig:ell}:} Contours of equal density for the joint slope and concavity distributions for two treatment groups that vary based on the magnitude of the biosignature 
$w = \bm{\alpha} \tran \bm{x}$ (left-to-right) and the angle $\theta$ between the biosignature covariate directions $\bm{\Gamma}_1$ and $\bm{\Gamma}_2$ {from model (\ref{lme})} (from top-to-bottom)}}
\label{fig:ell}
\end{figure}

The geometric illustrations are provided in Figure \ref{fig:ell}.
Let $w = \bm \alpha \tran \bm x$ denote the biosignature.   
The different panels in Figure \ref{fig:ell} depict several scenarios showing the separation between the distributions for the treatment groups based on two characteristics:
\begin{enumerate}
\item[(i)]
 $w= \bm \alpha \tran \bm x$, the magnitude of the biosignature which increases across the columns in Figure \ref{fig:ell} and 
 \item[(ii)] 
 $\theta=$ angle between the biosignature coefficient vectors $\bm{\Gamma}_1$ and $\bm{\Gamma}_2$ which increases across the rows in Figure \ref{fig:ell}.
\end{enumerate}

As the biosignature $w$ increases in magnitude, the two distributions move apart (along the green dotted lines) as can be seen going from left to right in the rows of Figure \ref{fig:ell} (unless the angle $\theta$ between the biosignature covariate directions $\bm \Gamma_1$ and $\bm \Gamma_2$ is zero as depicted in the panels in the top row of the figure).
Additionally, the larger the angle $\theta$ between the biosignature covariate directions $\bm \Gamma_1$ and $\bm \Gamma_2$, the quicker the two distributions separate as the biosignature $w$ increases in magnitude. 
Also note that the orientations of the equal contour ellipses (as determined by the eigenvectors of $\bm{D} _1$ and $\bm{D}_2$) impacts the degree of overlap between the two treatment distributions.
The situation when the biosignature is zero, $w=0$, for our illustration is shown in the very left column in Figure \ref{fig:ell} -- in this case, there is a large degree of overlap between two treatment distributions indicated by the large overlap in the equal contour ellipses.  
The goal of the TDR optimization is to determine $\bm{\alpha}$ that ``moves'' the two treatment distributions as far apart as possible based on these characteristics, that is, the goal is to maximize the longitudinal interaction effect.
 
\section{Simulation studies}\label{sim}

The performance of the TDRs based on determining a biosignature that maximizes the mean KLD is studied using simulation illustrations in this section.
Because one of the primary motivations for developing the approach is the EMBARC depression study, we use quadratic outcome trajectories for illustration and the 
simulation parameter settings are set to mimic the parameter estimates from the EMBARC study.
In all simulations, we consider the case with $K=2$ treatment groups.
The outcomes are generated from the model (\ref{lme}) using the quadratic functional approximation, i.e., setting $\bm g(t) = (1, t, t^2) \tran$.   
Therefore, for a subject with outcomes measured at time $\bm{t} = (t_1, ..., t_m)\tran$, the design matrix $\bm G(\bm t)$ has 3 columns: a column of ones, $\bm{t}$, and $(t^2_1, ..., t^2_m)\tran$.

\subsection{Longitudinal trajectory parameter settings}\label{long}

The {assessment time points are set to} $\bm t = (0,1,2,..., 7) \tran $, i.e., the outcomes are collected at baseline and subjects are followed up for 7 weeks. 
The fixed-effect coefficient parameters $\bm \beta_1$ and $\bm \beta_2$ are chosen to be similar to each other (as was the case {for the estimated} EMBARC study {parameters}), 
$$\bm{\beta}_1 = (20, 3, -0.5)\tran \;\;   \mbox{{and}} \;\;  \bm{\beta}_2 = (20,2.3,-0.4)\tran .$$
The vectors of random effects are set as $\bm{b}_{i,1} \sim MVN(\bm 0, \bm D_1)$ and $\bm b_{i,2} \sim MVN(\bm 0, \bm D_2)$ with 
$$\bm D_{1} =  \left(\begin{array}
{ccc}
0.5 & -0.1 & -0.01\\
-0.1 & 0.5 & -0.01 \\
-0.01 & -0.01 & 0.01
\end{array}\right) \;\; \mbox{and} \;\; \bm D_2 = \left(\begin{array}
{ccc}
0.4 & -0.12 & -0.01 \\
-0.12 & 0.5 & -0.01 \\
-0.01 & - 0.01 & 0.01
\end{array}\right)$$
The parameter $\bm \Gamma_k$ provides direction information, i.e., how the coefficients of the trajectories vary with changes of  the  baseline biosignature $\bm \alpha \tran \bm x$. 
For this simulation illustration, we set
$$\bm{\Gamma}_1 = (0, \cos(\theta), \sin(\theta))\tran, \bm \Gamma_2 = (0, \cos(\theta), - \sin(\theta))\tran$$
where $\theta$'s are chosen to be $0^{\circ}, 1^{\circ},  2^{\circ}, 5^{\circ}$ degrees, corresponding to the angles between vector $\bm \Gamma_1$ and $\bm \Gamma_2$ as $\theta = 0^{\circ}, 2^{\circ},  4^{\circ}, 10^{\circ}$. 
The error terms are simulated as $\bm{\epsilon}_{ik} \sim (N(0,1)$ and are all independent of each other. 
The average trajectories (with only fixed effects and without the effect of biosignatures) are illustrated in Figure \ref{fig:traj}:
\begin{figure}[H]
\caption{\textbf{Average Trajectories Plots for the Simulation}}
\centering
\includegraphics[width=0.65\textwidth]{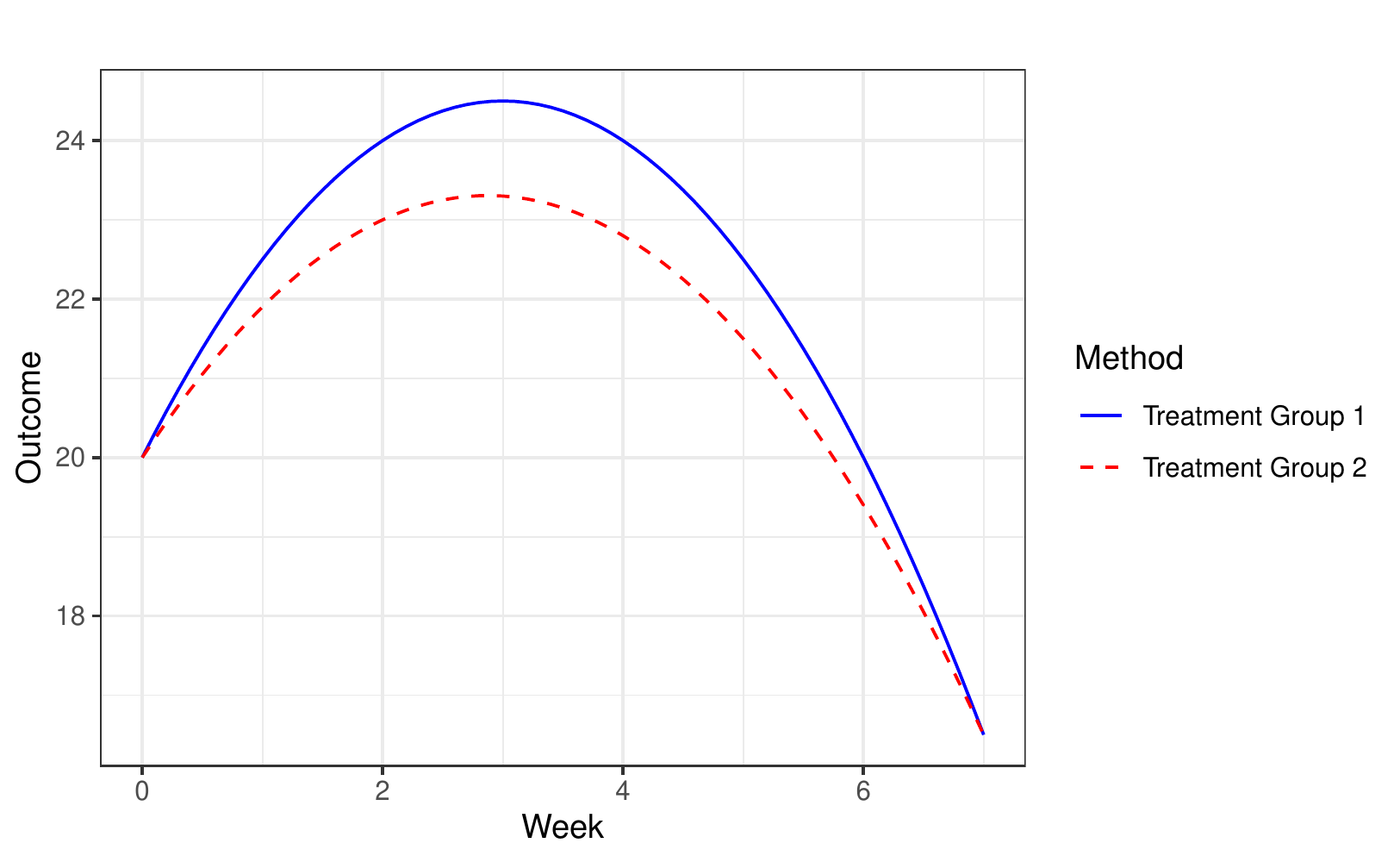}
\caption*{
\footnotesize{\textbf{Figure \ref{fig:traj}: } Trajectories from the treatment group 1 and 2 were generated based on their average trajectories, which are the blue solid and red dashed curves, respectively.}}
\label{fig:traj}
\end{figure}

From Figure \ref{fig:traj}, the mean curves of treatment group 1 and 2, without regard to the impact of baseline covariates, have similar shapes and they have exactly the same average tangent slope.  
Thus, based on the average trajectories, one would conclude there is not difference between the two groups.
However,  incorporating the patients'  baseline information via the biosignature, distinctions between the two groups arise. 
That is, tailoring treatment to individual patient using baseline characteristics, an optimal treatment decision is possible even though on average, subjects respond similarly.
        
        \subsection{Covariate parameter settings}\label{covariatesim}
        
        To generate the linear biosignature, we consider settings with the number of covariates $p$ to be $p = 2, 10, 20, 30$. We set $\bm \alpha = (1,..,p)\tran$, standardized to have norm one.  
        The baseline covariates $\bm x_{ik}$ were sampled from $MVN(\bm \mu_x, \bm \Sigma_x)$, where $\bm \Sigma_x$ is a $p \times p$ matrix and has 1's on the diagonal and $0.5^{|i-j|}$ for the element on the $i$th column and $j$th row. 
The vector of mean values, $\bm \mu_x$, is set as $\bm \mu_x = \big(-p, -(p-1),...,2,1\big)$, i.e., half are negative and the other half are positive, which makes the biosignature $\bm \alpha \tran \bm x$ have mean $0$. 

\subsection{Missing data settings}\label{missing}
Different scenarios of missingness are considered {in the simulations}: (i) no missing data, i.e., all measures from all subjects are collected; 
(ii) missing completely at random (MCAR), that is, for each subject, each of his/her outcomes at week $2, ..., 7$ has a probability of $40\%$ to be missing;
(iii) each patient has a $50\%$ chance of dropping out the trial at week $2$ (Dropout). Note that the outcomes at baseline are {recorded}  for all patients.  

\subsection{TDRs and simulation settings }\label{tdrsims}

The proposed TDR based on a longitudinal single-index model to maximize the Kullback-Leibler Divergence (denoted as LS-KLD) is compared with a three other approaches for constructing TDRs The TDR methods considered are:
\begin{enumerate}
\item[(i)] Longitudinal single-index model based on Kullback-Leibler Divergence (denoted as LS-KLD),
\item[(ii)] SIMML \citep{park2020single} estimated from maximizing the profile likelihood,
\item[(iii)] Linear GEM model \citep{petkova2017generated} estimated under the criterion of maximizing the difference in the treatment-specific slopes, denoted as linGEM,
\item[(iv)] The outcome weighted learning (OWL) method~\citep{zhao2012estimating} based on a Gaussian radial basis function kernel. 
\end{enumerate}
Note that the LS-KLD method estimates the TDR  (\ref{tdr}), based on the ATS estimator, while the SIMML and linGEM assess the TDR with the differences in the observed outcomes from baseline to last observation (change scores).
Additionally, the TDRs estimated by plugging the actual $\bm \alpha$ in (\ref{tdr}) (instead of the estimated $\bm{\alpha}$) are also presented for comparison.
For each scenario, a training data set with sample size $n = 200$ ($100$ per group) {was used to estimate the model parameters and the TDRs}.   
Testing data sets with $n = 1000$ were generated to obtain an independent assessment of the performances of each of the TDRs.
For the $i$th simulated subject in the testing data, the true treatment assignment was calculated using
\begin{equation}\label{tdr-test}
\mathcal{D}_0(\bm x) = I \Big ( \frac{\bm g(t_m^*)\tran - \bm g(t_1^*)\tran}{t_m^* - t_1^*} \big({\bm \beta}_2 + \bm b_{i2} + {\bm \Gamma}_2 ({\bm \alpha} \tran \bm x )\big) > \frac{\bm g(t_m^*)\tran - \bm g(t_1^*)\tran}{t_m^* - t_1^*} \big({\bm \beta}_1 + \bm b_{i1} + {\bm \Gamma}_1 ({\bm \alpha} \tran \bm x) \big)\Big) + 1.
\end{equation}
Note that the random effects $\bm b_{i1}, \bm b_{i2}$ were considered when determining the true group label since the random effects are available in the simulations, i.e. the true trajectories for each subject were known.
The estimated group assignment was evaluated with $\mathcal{D}(\bm x)$ in (\ref{tdr}). 
The performances of models were assessed with Proportion of Correct Decision (PCD), as 
\begin{equation} \label{pcd0}
    \text{PCD: }\text{ }\frac{1}{n} \sum_{i=1}^n I \Big( \mathcal{D}(\bm x_i) = \mathcal{D}_0(\bm x_i)\Big).
\end{equation}
The simulations were conducted in R 3.5.1 \citep{rrr}. The mixed-effect models were fitted with the package lme4 and the optimization {to estimate} $\bm \alpha$ is {performed using the} optim function in R. 
Altogether, this simulation experiment is examining  $3 \times 4 \times 4 = 48$  different scenarios: three missing  data scenarios, four different dimensions for the baseline features, and four angles between $\bm \Gamma_1$ and $\bm \Gamma_2$. 200 repetitions were conducted for each scenario.
 
 \subsection{Simulation results}\label{simresults}
 
In order to compare the different TDRs, the proportion of correct decisions (PCD) of the test sample that is correctly classified to the optimal treatment is computed following (\ref{pcd0}). 
That is, each subject in the test sample is simulated with known labels of group assignments. A correct decision is defined as the match between the estimated and true group assignment. 

\begin{figure}[H]
\caption{\textbf{Proportion of Correct Decisions (PCD)}}
\centering
\includegraphics[width=1\textwidth]{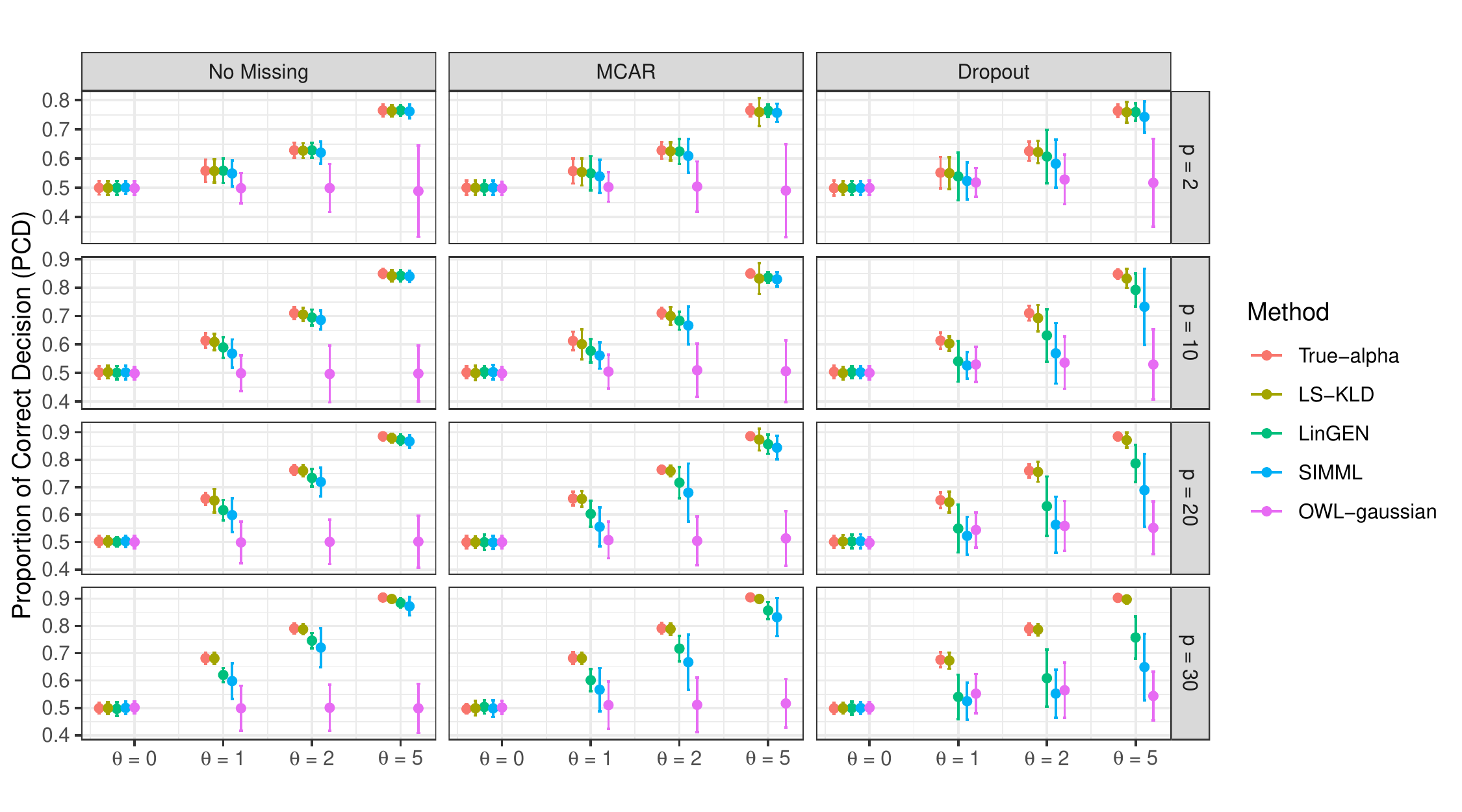}
\caption*{
\footnotesize{\textbf{Figure \ref{fig:pcd}: }Plots of the proportion of correct decisions (PCD) of the treatment decision rules obtained from 200 training data sets fro each of the four methods. Each panel corresponds to one of the combinations of $p \in \{2, 10, 20, 30\}$ and no missing data, MCAR, or Dropout. The dots represent the mean PCD estimated across the 200 repetitions and the bars represent the 1.96 standard deviations. }}
\label{fig:pcd}
\end{figure}

Figure \ref{fig:pcd} summarizes the results {for all} simulation scenarios. 
From left to right, the columns present the scenarios of no missing data, MCAR, and dropout.
The panels across the rows show the PCD estimations under different number of covariates, $p$. 
The mean PCDs with 1.96 standard deviations are plotted.
From left to the right, the bars present the PCD estimated using the actual $\bm \alpha$, the estimated longitudinal single-index model (LS-KLD), SIMML, linGEM and outcome weighted learning with Gaussian kernal (OWL-gaussian).
For each panel, the $y$-axis gives the PCD while the $x$-axis represents the angle  $\theta$ between $\bm \Gamma_1$ and $\bm \Gamma_2$.
When $\theta = 0^{\circ}$, all methods have PCD around $50\%$ across all scenarios as expected, i.e. the methods assign subjects to group 1 or group 2 essentially by chance (which is expected with $\theta=0$).
For each panel, as $\theta$ increases (left to right), the separation of the groups increases and all methods perform better instead of OWL-gaussian. 
The performances in terms of different $\theta$ can be explained through the illustration in Figure \ref{fig:ell}. 
When $\theta = 0^{\circ}$ and the ellipses shift in the same direction in which case the biosignature is not able to separate the groups; 
the opposite extreme is when $\theta = 180^{\circ}$ corresponding to maximizing the separation between the treatment group distributions where the two distributions move in opposite directions as the biosignature varies.
Across all the scenarios, the LS-KLD model has similar performance as using the true $\bm \alpha$ to estimate the PCD. 
When there is no missing data, the TDRs estimated by LS-KLD, SIMML, and linGEM perform similarly when the number of covariates is small (e.g. $p = 2$). The OWL-gaussian is outperformed by the other three methods. When $p$ gets larger, the LS-KLD shows larger PCD values compared to SIMML and linGEM, especially when $\theta = 1^{\circ}$ or $2^{\circ}$. 
On the other hand, when there is missing data, the LS-KLD method performs much better than the other approaches, in terms of higher PCDs. 
The outcome weighted learning approach has bad performance across all scenarios. 
This may be caused by the weakness of this algorithm. The outcome weighted learning, which optimizes treatment decisions based on weighted support vector machine (SVM), is not robust against a perturbation of the clinical outcome.
For example, it could break down by a simple shift of a clinical outcome~\citep{zhou2017residual}. 

In addition to $\theta$, the speed of separation of the two distributions is also determined by the norms of $\bm{\Gamma}_1$ and $\bm{\Gamma}_2$. 
By applying the proposed optimization procedure for LS-KLD, estimates $\hat{\bm \alpha}$ are obtained that maximally separate the treatment groups for an optimally identified biosignature $\hat{\bm \alpha}\tran \bm x$. 
The resulting TDR then  provides high confidence in assigning subjects to treatment groups among subjects who have a relatively high magnitude of $\hat{\bm \alpha}\tran \bm x$.

\section{Application to the EMBARC study}\label{sec:embarc}

\begin{figure}[H]
\caption{\textbf{Treatment-specific Spline Approximated Regression Curves}}
\label{fig:spline}
\centering
\includegraphics[width=0.8\textwidth]{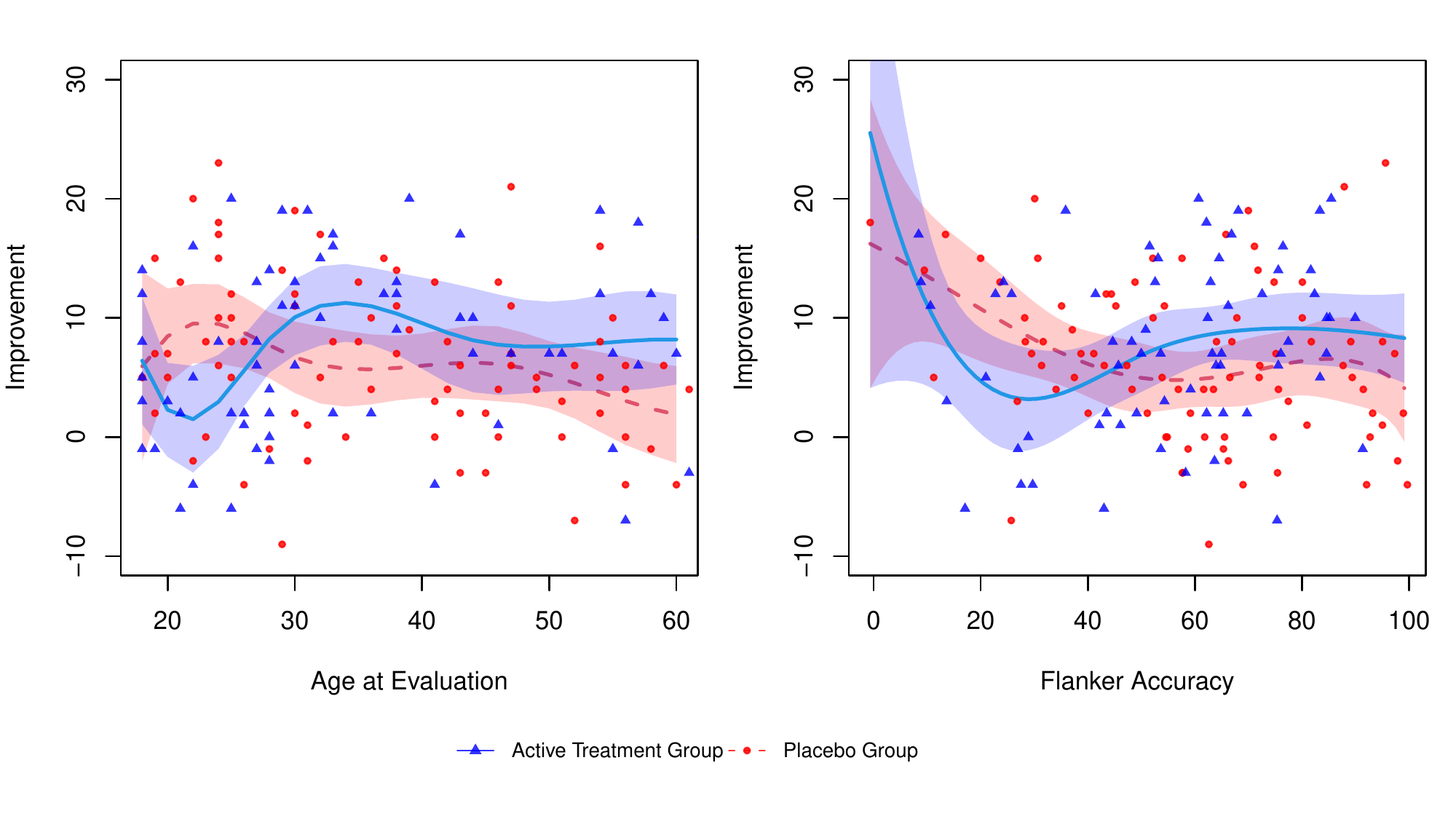}
\caption*{
\footnotesize{
\textbf{Figure \ref{fig:spline}:} The scatter plots present the relationship between the outcome improvement and the baseline covariates, ``Age at Evaluation" (left panel) and ``Flanker Accuracy" (right panel). 
The placebo and active treatment groups are shown by the red round and blue triangle points, respectively.
The data points are overlaid with treatment-specific spline approximation spline regression curves with 5 degrees of freedom. 
The red dashed curves represent the placebo group, whereas the blue solid curves represent the active treatment group.}}
\end{figure}

To illustrate the performance of the LS-KLD method to identify a linear combination of baseline covariates to define a TDR in the EMBARC study (see Section \ref{embarcSection}), 
10 baseline covariates $\bm X = (x_1, ..., x_{10})\tran$ from the the Demographic, Clinical Measurement and Behavioral Phenotyping modality are used. 
The outcomes $Y$ are the HDRS measured at week 0 (baseline), and weeks 1, 2, 3, 4, 6, and 8.
Of 160 the patients who had data on these covariates, 87 were randomized to placebo and 73 to the antidepressant group -- Table \ref{tab2:} summarizes these varaiables by treatment group.
Summary statistics for the active treatment and placebo arms (e.g. mean and standard deviations for continuous covariates; counts and percentages for categorical covariates) are shown in the second and third columns, respectively.
Linear regression models with the change scores (improvement in HDRS from week 0 to last week) as the outcome variable and the group indicator, the baseline covariate and their interaction as the predictor variables were fitted for each baseline covariate. 
The significance of the interaction terms were investigated using a likelihood ratio test.
In Table \ref{tab2:}, the $p$-values of LRTs are summarized in column 4.
Treatment had significant interaction effects for two baseline factors, ``Age at Evaluation" and ``Flanker Accuracy".
\begin{center}
\begin{table*}[t]%
\caption{Demographic Table.\label{tab2:}}
\centering
\begin{tabular*}{430pt}{p{6cm}ccccc}
\toprule
&\multicolumn{2}{c}{\textbf{Mean (SD) / N ($\%$)}} & & \multicolumn{2}{c}{IPWE Value} \\
\cmidrule{2-3}\cmidrule{5-6}
\textbf{Baseline Covariates} & \textbf{Treatment}  & \textbf{Placebo}  & \multicolumn{1}{@{}l@{}}{\textbf{$p-$value}}  & \textbf{Linear}& \textbf{Nonpar.}   \\
\midrule
$(x_1)$ Age at Evaluation & 37.44 (15.01) & 38.30 (13.03) & 0.006 & 5.71 & 5.13\\
$(x_2)$ Anger attack & 3.08 (2.16) & 2.97 (2.07) & 0.839 & 6.66 & 7.23\\
$(x_3)$ Axis II & 3.92 (1.40) & 3.91 (1.50) & 0.410 & 6.56 & 7.20 \\
$(x_4)$ Chronicity & & &  0.219 & 7.08 & 7.08 \\
$~$ $~$ $~$ $~$ $~$ $~$ Yes & 39 (53.4) & 47 (54.0) &  &  & \\
$~$ $~$ $~$ $~$ $~$ $~$ No & 34 (46.6) & 40 (46.0) & & & \\
$(x_5)$ Hypersomnia & & & 0.667 & 6.33 & 6.33 \\
$~$ $~$ $~$ $~$ $~$ $~$ Yes & 17 (23.3) & 16 (18.4) & & &  \\
$~$ $~$ $~$ $~$ $~$ $~$ No & 56 (76.7) & 71 (81.6)  &  &  &\\
$(x_6)$ Sex & & & 0.921 & 6.36 & 6.36\\
$~$ $~$ $~$ $~$ $~$ $~$  Female & 26 (35.6) & 32 (36.8) & & & \\
$~$ $~$ $~$ $~$ $~$ $~$ Male & 47 (64.4) & 55 (63.2) &  &  & \\
$(x_7)$ Number of correct responses  & 0.05 (0.82) & 0.23 (0.71) & 0.309 & 6.46 & 6.55\\
$(x_8)$ Median Reaction time & 0.51 (1.78) & 0.01 (1.10) & 0.649 & 6.60 & 6.27\\
$(x_9)$ Flanker Reaction Time & 61.36 (28.61) & 59.52 (24.40) & 0.075 & 5.65 & 5.77\\
$(x_{10})$ Flanker Accuracy & 0.22 (0.15) & 0.22 (0.15) & 0.005 & 5.44 & 6.00 \\
\bottomrule
\end{tabular*}
\begin{tablenotes}
\item \textbf{Table \ref{tab2:}:} The second and third columns summarize the mean (SD) for continuous variables and the counts ($\%$) for categorical variables in each group.
``Number of correct responses" represents the number of correct responses in the ``A not B” Working memory task.
The $p$-values for the interaction effects between outcomes and each covariate are provided in the fourth column.
LRTs were used to calculate the $p-$values. 
The IPWE values obtained by fitting linear regressions or B-spline regressions are shown in the last two columns.
\end{tablenotes}
\end{table*}
\end{center}
To highlight the interaction effects of age at evaluation and flanker accuracy with treatment from Table \ref{tab2:}, Figure \ref{fig:spline} shows
cubic B-spline regressions (with five degrees of freedom) fitted separately for the placebo (red dashed lines) and antidepressant treatment (blue solid lines) groups, with the 95$\%$ confidence bounds (represented as the red and blue areas in the plots).
Despite evidence of the interactions between the curves in Figure \ref{fig:spline}, the corresponding confidence intervals overlap substantially indicating that these
two covariates individually have limited treatment-modifying effects and TDRs based on these variables individually may perform poorly.

In order to compare the performance of different TDRs, the {\em value} of the TDR will be used. 
The value of a TDR is defined to be the average outcome when treatments are assigned based on the TDR.
We shall evaluate the effectiveness of a TDR using an empirical estimate of the value defined as the inverse probability weighted estimator (IPWE) ~\citep{murphy2005generalization}:
\begin{equation}\label{ipwe}
    \text{IPWE} = \sum_{i = 1}^n \{I(A_i = \hat A_i) U_i \}/ \sum_{i = 1}^n I(A_i = \hat A_i),
\end{equation}
where
$A_i \in \{1,2\}$ is the observed treatment assignment and $\hat{A}_i$ is the treatment assignment based on the estimated TDR  (1 = active treatment, 2 = placebo);
$U_i$ is the change score, i.e., the difference between the first and last HDRS outcomes. 
Change score is being used here to compare TDRs since this is a measure that is extractable from all methods under consideration.
Larger values of $U_i$ are indicative of an effective treatment (more improvement from baseline).

The treatment modifying effect for each baseline characteristic was evaluated with the IPWE (\ref{ipwe}). 
Simple linear regressions and $B$-spline regressions of the HDRS improvement on each covariate were obtained separately in the antidepressant and the placebo groups. 
A 10-fold cross-validation (CV) was conducted to assess the TDR performances and then this 10-fold CV was repeated for 100 random splits of the data and then averaged over all random spits.       
The CV from applying linear regressions (Linear) and B-spline regressions (Nonpar.) are summarized in the fifth and sixth column in Table \ref{tab2:}. 
The IPWEs estimated with the $B$-spline regressions have slightly larger values than the linear regression. 
Additionally, an all-drug (i.e., a ``one-size-fits-all'') policy was also applied based on the decision rule to give everyone the active treatment resulting in an IPWE  $= 7.49$. 
Thus, the policy to treat everyone with the active drug performs better than either regression approach (linear or $B$-spline) when considering a single predictor variable.
Although the two baseline covariates, ``Age at Evaluation" and ``Flanker Accuracy", have significant interaction effects with the treatment, they perform poorly in terms of informing an optimal treatment assignment.
        
Since no single predictor appears to be strong effect modifier by itself, results for the single-index models with a combination of all the baseline covariates were examined based on the change score.
The four TDRs used in the simulation study in Section \ref{sim} were considered: LS-KLD, SIMML, linGEM and OWL-gaussian. 
These TDRs were compared using 10-fold CV with 100 random splits as was done for models with a single predictor.
Results are also presented based on assigning all subjects to the active treatment group (AllDrug) or assigning all subjects to placebo (AllPbo).
        
        \begin{figure}[H]
        \caption{\textbf{Evaluation of the IPWE}}
        \label{fig:box}
        \centering
        \includegraphics[width=0.8\textwidth]{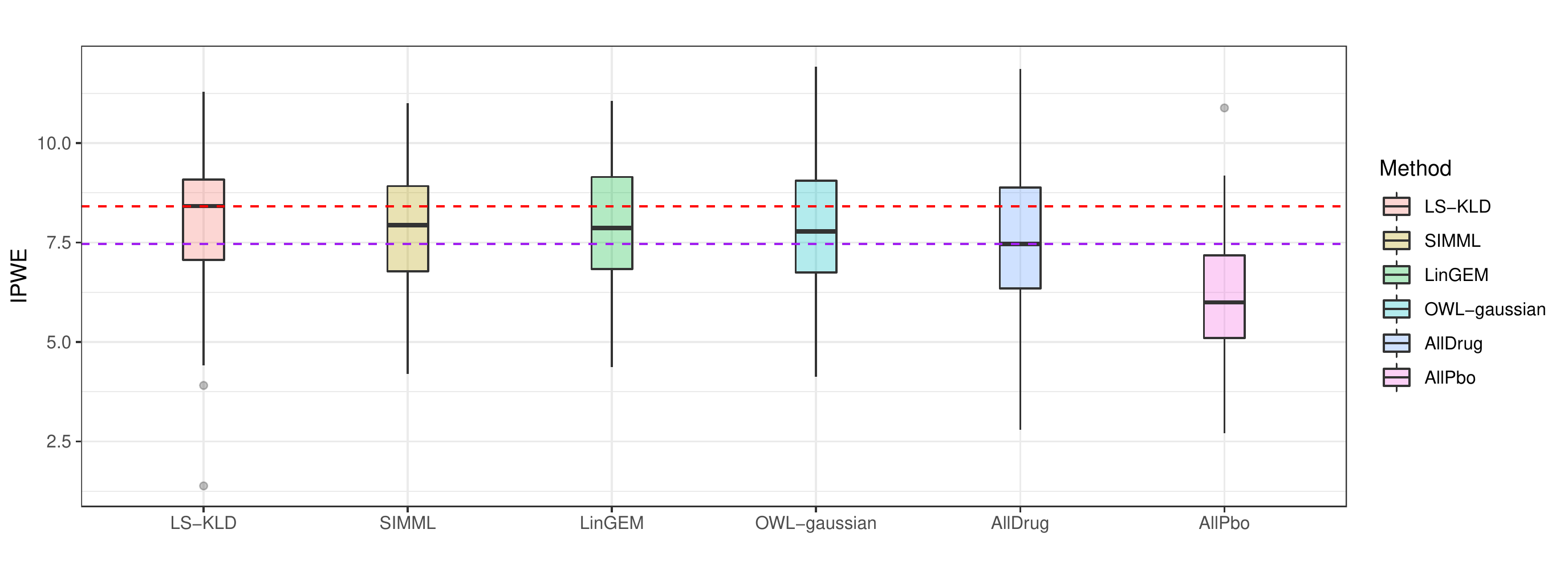}
        \caption*{
          \footnotesize{\textbf{Figure \ref{fig:box}: }
            Box plots of the estimated IPWE values across the 1000 splits (higher values are preferred). The mean and standard deviations for each method are: LS-KLD: 8.41 (1.69); SIMML: 7.94 (1.48): LinGEM: 7.87 (1.58);
            OWL-gaussian: 7.84 (1.67); AllDrug: 7.47 (1.7); AllPbo 6.00 (1.52) .
          }}
        \end{figure}
        
The IPWEs for the TDRs are summarized in Figure \ref{fig:box}. 
The purple horizontal dashed line shows the median IPWE value across the 1000 repetitions (10-fold times 100 random splits) when all patients are assigned to the treatment group, while the red dashed line represents the  median IPWE value of the {LS-KLD}.
All single-index models have higher median IPWE values than the ``one-size-fits-all' approaches (AllDrug, AllPbo), implying that a treatment decision rule based on an individual's biosignature can achieve better performance than the policy of administering the active treatment to everyone.
The IPWE estimations of SIMML, LinGEM and OWL-gaussian are fairly similar, but their performance is worse than the LS-KLD approach.
        
\begin{figure}[H]
\caption{\textbf{Approximated Spline Regression with Biosignatures}}
\label{fig:single}
\centering
\includegraphics[width=0.8\textwidth]{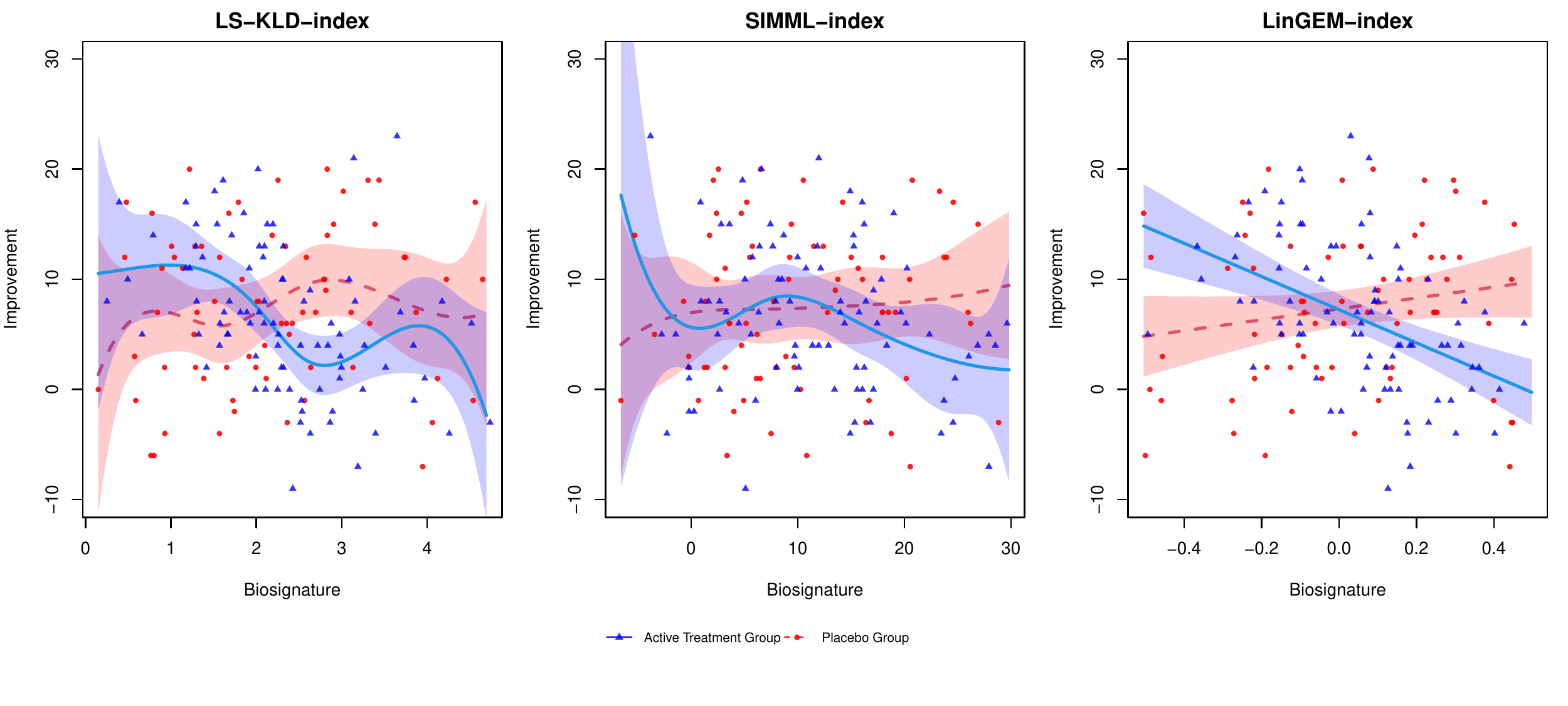}
\caption*{
\footnotesize{\textbf{Figure 7:} The scatter plots present the relationship between the outcome improvement and the estimated biosignatures by each single index model. The placebo and treatment groups are shown by the red and blue points, respectively. The data points are overlaid with treatment-specific spline approximation regression curves with 5 degrees of freedom. The red curves represent the placebo group, whereas the blue curves represent the treatment group.}}
\end{figure}

The single index $\hat{\bm \alpha} \tran \bm X_i$ was calculated for each subject $i$ with the coefficients estimated from the three single-index models. 
The relationship between the outcome improvement and the single-index values are shown in Figure \ref{fig:single}.
Red round and blue triangle points represent the outcomes from the placebo and active treatment groups respectively.
Cubic $B$-spline regressions were fitted separately for the placebo (red dashed curves) and active treatment (blue solid curves) groups, with the 95$\%$ confidence bands represented as red and blue areas in the plot.

\subsection{Additional TDR Comparisons}\label{additional}

The above results show that the LS-KLD approach shows the best in terms performed very well in terms of IPWE with the 10  baseline covariates that were selected for the illustration.
However, the performances of the TDRs depend on the choice of the predictors included in the model.  
The results for different TDRs can vary substantially depending on which covariates are used in the TDR. 
In order to obtain a more robust picture of how different TDRs perform  generally, we examined TDR performance for 500 models with various combinations of baseline covariates where the 500 models include random combinations of predictors with varying numbers of predictors.
The demographic, clinical, and behavior phenotyping data sets contained 21 covariates at the baseline. 
For the $i$th combination, the number of covariates $n_{ci}$ was chosen at random from a uniform distribution on the set $\{8,9,\dots, 21\}$. Then $n_{ci}$ covariates were randomly selected from the demographic, clinical, and behavior phenotyping covariates set.
Since the ``Age at Evaluation" and ``Flanker Accuracy" have significant interactions with the outcome, we ensured each covariate combination for chosen models contain these two covariates, as well as the ``Flanker Reaction Time", which is an  {important}  measure in clinical practice ~\citep{dillon2015computational}.

\begin{figure}[H]
        \caption{\textbf{Density Plots of the IPWEs}}
        \label{fig:density}
        \centering
        \includegraphics[width=0.6\textwidth]{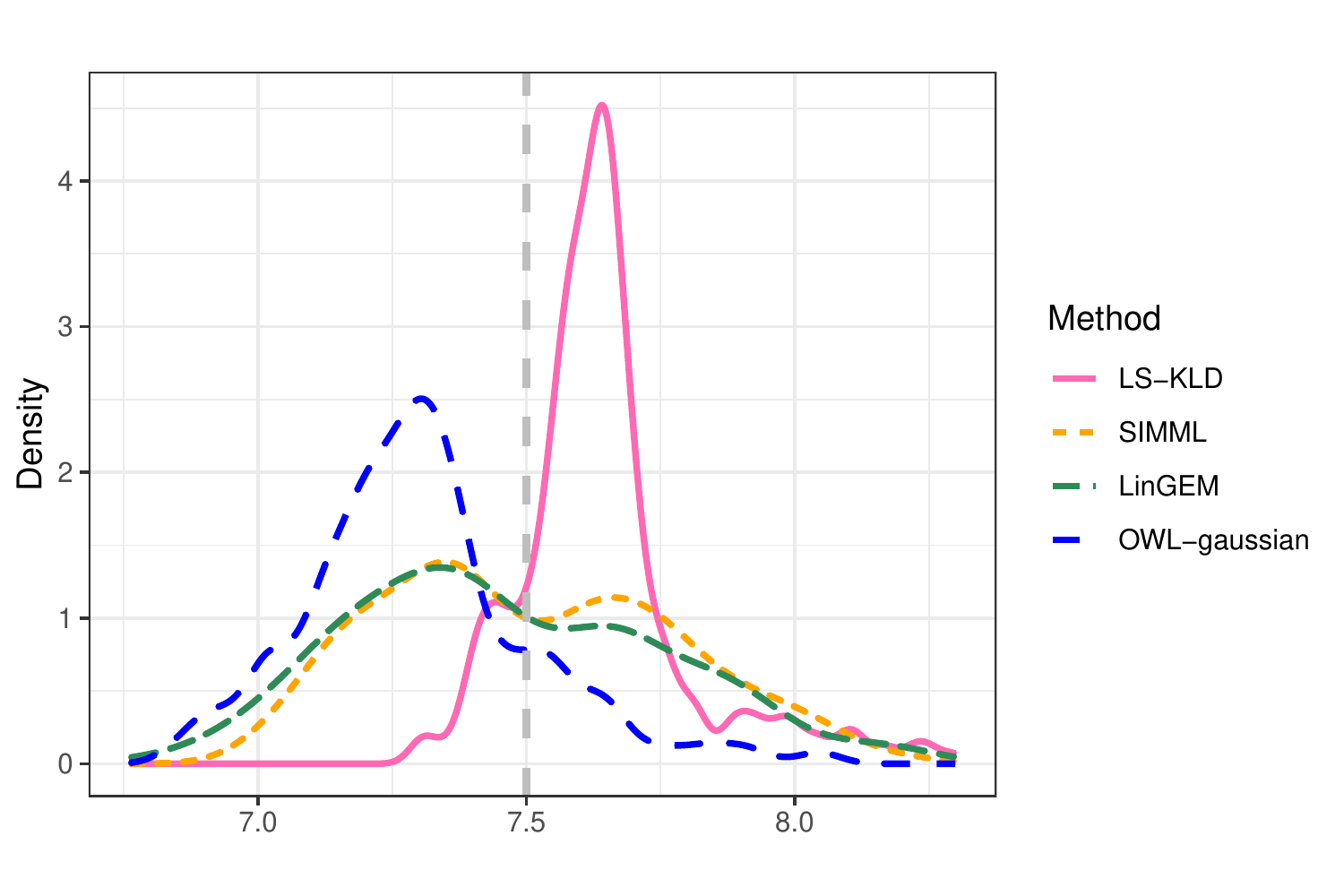}
        \caption*{
          \footnotesize{
            \textbf{Figure \ref{fig:density}: } The density plots of the mean {CV} IPWE values across 500 combination of baseline covariates. The vertical dashed line {re}presents the IPWE value by assigning all subjects to the drug group. The mean (SD) of each method is: LS-KLD: 7.64 (0.18); SIMML: 7.38 (0.33); LinGEM: 7.32 (0.35)
            ; OWL-gaussian: 7.30 (0.21).}}
        \end{figure}

100 repetitions of 10-fold CV were conducted and the mean IPWE values were calculated across the 1000 training sets. 
Density plots of the IPWEs estimated by TDR approaches (LS-KLD, SIMML, linGEM, OWL-gaussian) across the 500 baseline covariate combinations are shown in Figure \ref{fig:density}.
 By simply allocating all individuals to the active treatment group (AllDrug), the IPWE value $= 7.49$ (shown as the vertical dashed line).
The density plots of LS-KLD, SIMML, linGEM, and OWL-gaussian are represented by the pink, yellow, green and blue curves, respectively.
OWL-gaussian performs worst with mean IPWE $=7.30$, and a relatively large standard deviation 0.21.
The SIMML and linGEM perform similarly in terms of IPWE estimation.
For SIMML, the mean (SD) IPWE value across the 500 combinations is 7.38 (0.33), while for the linGEM model, it is 7.32 (0.35).
SIMML has slightly better performance than {the} linGEM, in terms of a greater mean value of IPWE and a smaller standard deviation. 
However, those three mean values are smaller than the IPWE calculated by simply assigning all subjects to the active treatment group:
SIMML and linGEM only perform better than the policy of AllDrug 35$\%$ and 28$\%$ times , respectively. 
On the other hand, the mean IPWE is 7.64 and 82$\%$ combinations have better performance than Alldrug; additionally, the IPWE values estimated by LS-KLD have much smaller standard deviation (0.18) compared with the other two methods.  

Note that in this analysis, there are several examples of combinations of variables where the TDRs based on a single-index perform worse than the AllDrug policy.  
This is expected because some of the models may include variables that have negligible moderator effects and no combination of these variables may yield a strong TDR.
The LS-KLD method, that takes into account the longitudinal structure of the data, performs considerably better compared to SIMML, linGEM, and OWL-gaussian and also performs better than the AllDrug policy in the majority of scenarios. Besides, we included “Age at Evaluation", “Flanker Reaction Time", and “Flanker Accuracy" in all combinations.  Similar results are obtained without this restriction on containing these 3 covariates.

\section{Discussion}\label{disc}
        
In randomized clinical trials, patient heterogeneity is often a major concern, rendering a universal treatment strategy ineffective. 
Another significant issue in medical research is the problem of missing data.
Most randomized trials are longitudinal in nature but most precision medicine approaches to develop TDRs ignore the longitudinal structure of data.  
The TDR approach developed and illustrated in this paper is based on the longitudinal structure of the data which, in addition to making more robust use of the data, also allows for a more powerful approach to deal with missing data compared to other popular TDRs.
        
Another feature of the LS-KLD treatment decision rule developed in this paper is that the optimal single-index biosignature (i.e., linear combination of baseline features) is determined based on the average tangent slope which is an overall measure of improvement (or deterioration) across the longitudinal trajectory.
The estimator of the average tangent slope is more efficient and robust to missing data (e.g. MCAR) than other scalar summary measures (e.g. score change between last and first observation). Hence, the LS-KLD method has better performance, e.g., a higher proportion of correct decisions, when there is missing data.
        
There are several avenues for future research on the use of a Kullback-Leibler divergence to define treatment decision rules. For example, future work will focus on developing efficient and faster algorithms for implementing the LS-KLD TDR with variable selection.  
Extensions will also investigate the use of more flexible nonparametric link functions, i.e., consider a smooth function $h$ of the biosignature $h(\bm \alpha \tran \bm x)$ in (\ref{lme}), as is done in the SIMML model \citep{park2020single} and incorporating functional predictors into the 
TDR \citep[e.g.,][]{Ciarleglio.2015, ParkEtAl2021}.  
Further investigations into the impact of missing data under different missing data mechanisms (e.g., missing not a random) is currently being studied.

\bibliographystyle{unsrtnat}

\end{document}